\documentclass{phbauth}        % Don't modify
\usepackage{graphicx}

\begin{document}

\begin{frontmatter}

% Use lower case letters in the title.
\title{
  One-electron spectral functions of the attractive Hubbard model at
intermediate coupling.
}

\author[address1]{M.Yu.Kagan\thanksref{thank1}},
\author[address2]{R.Fr\'{e}sard},
\author[address3]{M.Capezzali},
\author[address2]{H.Beck},

\address[address1]{
    P.L.Kapitza Institute for Physical Problems, Moscow
117334, Russia
}
\address[address2]{
    Institut de Physique, Universit\'{e} de Neuch\^{a}tel, 2000 Neuch\^{a}tel,
    Switzerland
}
\address[address3]{
Department of Physics, Queen's
University, Kingston, ON, K7L 3N6, Canada
}

% The corresponding author should be distinguished and his email
% address and/or fax number must be given. His mailing address has to
% be complete: the proofs are send to this address around
% January 1, 2000. The address for sending proofs has to be indicated
% as "present address", if it is different from the address above.
\thanks[thank1]{Corresponding author. Present address:
    P.L.Kapitza Institute for Physical Problems, Kosysin str. 2, Moscow
117334, Russia.
E-mail: kagan@kapitza.ras.ru}

\begin{abstract}
  We calculate the one-electron spectral function
of the attractive-$U$ Hubbard model in two dimensions. We work in the
intermediate coupling and low density regime and evaluate
analytically the self-energy. The results are obtained in a
framework based on the self-consistent $T$-matrix
approximation. We also calculate the chemical potential of
the bound pairs as a function of temperature. On the basis
of this calculation we analyze the low-temperature
resistivity and specific heat in the normal state of this
system. We compare our results with recent beautiful
tunneling experiments in the underdoped regime of
HTSC-materials.
\end{abstract}


\end{frontmatter}

% The main text begins here. The \section commands are optional.
\section{Introduction}
The attractive Hubbard model is
%one of the basic models for studying
%superconductivity. Despite its simplicity, this model is
very challenging for
theorists since its physics is bearing quite some resemblance to
the underdoped regime
in High Temperature Superconductors \cite{1}, \cite{2}.

In the present paper we treat analytically the intermediate coupling regime of
this model.
%, where we can expect a nontrivial interplay between the
%quasiparticles and the bound pairs.
There are two characteristic
temperatures in
this regime. The first one is a crossover temperature $T_\ast$, where
the density
of electrons in bound pairs $2n_{\mathrm{B}}$ is equal to the density
of unpaired
electrons $n_F$. Of course, the total density of charge carriers in the system
satisfies the condition $n=n_F+2n_{\mathrm{B}}$. The second temperature
$T_{\mathrm{KT}}$ is a Kosterlitz-Thouless critical temperature \cite{3} of a
superfluid transition in 2D. We consider the low-temperature normal
regime of the
model $T_{\mathrm{KT}}<T<T_\ast$.
%In this regime the Saha formula is valid
%\cite{4}. It describes the kinetical equilibrium in the ionized gas and yields
%$n_{\mathrm{B}}/n^2_F \sim e^{T_\ast/T}$ .

%\section{Theoretical model}
\section{Theoretical model and results}
%We study the Hubbard model on the square lattice:
%\begin{equation}
%\hat{H}=-t \sum_{\langle ij\rangle} c^{\dagger}_{i,\sigma} c_{j,\sigma}
%+ U \sum_{i} n_{i,\uparrow} n_{i,\downarrow}.
%\end{equation}
We consider an attractive-$U$ Hubbard model \cite{5}
%interaction
($U<0$) in the
intermediate coupling regime $|U|{\ \lower-1.2pt\vbox{\hbox{\rlap{$ <
$}\lower5pt\vbox{\hbox{$\sim$}}}}\ }W $.
%,
%where $W$ is the bandwidth.
We restrict our calculations to low density
limit
%of
%charge carriers $n<<1$.
In this limit the crossover temperature
$T_\ast$ is given
by \cite{5}:
%{\large
\begin{equation}
T_\ast=\frac{|E_{\mathrm{b}}|}{2\ln( {|E_{\mathrm{b}}|}/{Wn})},
\end{equation}
%}
where $|E_{\mathrm{b}}|$
%{\large $\frac{W}{\left[
%\exp\frac{W}{|U|}-1\right]}$}$<W$
is the binding energy of a local pair on the
empty lattice \cite{5}.
At the same time, the Kosterlitz-Thouless transition temperature reads \cite{6},
\cite{7}:
\begin{equation}
T_{\mathrm{KT}}=\frac{nW}{4\ln \ln{ (|E_{\mathrm{b}}|}/{Wn})}.
\end{equation}
Low density means that $T_{\mathrm{KT}}<T_\ast$, or, in another words:
$|E_{\mathrm{b}}|>${\large $\frac{nW}{2}$}.

%\section{Results}
%We work in the $T$-matrix approximation. Our aim is to find the contribution of
%two-particle bound states to the one-particle spectral function.
We
first calculate
the two-particle $T$-matrix for small $\omega$ and $q$. This yields:
\begin{equation}
T(\omega,{\bf q}) = \frac{|E_{\mathrm{b}}|W}{\left( \omega -\frac{q^2}{4m}
+\mu_{\mathrm{B}} +i0 \right)}. \end{equation}
It is important to note that the pole-structure of the
$T$-matrix reflects the creation of a bound pair with a mass
$m_{\mathrm{B}}=2m=1/t$ and a bosonic chemical potential
$\mu_{\mathrm{B}}=2\mu + |E_{\mathrm{b}}|$, where $\mu$ is the
one-particle chemical potential.
%The pole of the $T$-matrix
%gives the following contribution to the one-particle self
%energy:
%\begin{equation}
%\Sigma=\frac{|E_{\mathrm{b}}| Wn}{(\omega
%+\varepsilon_{\mathrm{q}}-\mu+\mu_{\mathrm{B}} +i0)}, \end{equation} where
%$\varepsilon_{\mathrm{q}} = q^2/2m$ is the one-particle
%spectrum.
As a result, we obtain the following expression
for the one-particle spectral function:
\begin{eqnarray}
\hbox{Im} G(\omega, {\bf q})& =& \left[ 1-\frac{|E_{\mathrm{b}}|
Wn}{(\varepsilon_{\mathrm{q}}-\mu)^2} \right] \delta
(\omega - \varepsilon_{\mathrm{q}}+\mu) \nonumber \\
&+& \frac{|E_{\mathrm{b}}| Wn}{(\varepsilon_{\mathrm{q}}-\mu)^2} \delta
(\omega + \varepsilon_{\mathrm{q}}-\mu+\mu_{\mathrm{B}}).
\end{eqnarray}
This expression describes an assymmetric two-band structure, consisting of a
fermionic band and a bosonic band separated by a correlation gap
$\Delta=|E_\mathrm{b}|$.
%with a particle-like dispersion
%$\omega_{\mathrm{F}}=\varepsilon_{\mathrm{q}}-\mu$
%and a bosonic band with a
%hole-like dispersion $\omega_{\mathrm{B}}=\mu-
%\mu_{\mathrm{B}}-\varepsilon_{\mathrm{q}}$. The density of states in
%the fermionic
%band is equal to
% $N_\mathrm{F}(\omega)=${\large $\frac{1}{W}$}$\left(
% 1-\right. ${\large $\frac{|E_{\mathrm{b}}| Wn}{\omega^2}$}$\left. \right)$
%$N_{\mathrm{F}}(\omega)= \frac{1}{W} \left(
%1- \frac{|E_{\mathrm{b}}| Wn}{\omega^2} \right)$
%for frequencies {\large $\frac{|E_{\mathrm{b}}|}{2}$}$ <\omega <W$. The
%density of states in the bosonic band is equal to
%$N_{\mathrm{B}}(\omega)= \frac{1}{W} \frac{|E_{\mathrm{b}}| Wn}{\omega^2}$
%for frequencies $-W< \omega<${\large $-\frac{|E_{\mathrm{b}}|}{2}$}. This
%structure of the density of states reflects the appearance of a
%correlation gap
%$\Delta=|E_{\mathrm{b}}|$ between the bands.
For $T_{\mathrm{KT}} < T <
T_{\ast}$:
%this gap is
%quite stable, leading to an exponentially small number of particles
$n_{\mathrm{F}} \sim \exp \left\{ -\frac{|E_{\mathrm{b}}|}{2T}
\right\}$,
%in the fermionic band.
%
%Of course, after frequency integration of
%$N_{\mathrm{F}}(\omega)$ and $N_{\mathrm{B}}(\omega)$ we will
%restore the requirement of particle conservation:
%$n=n_{\mathrm{F}} +2n_{\mathrm{B}}$.
%and
hence
%for $T_{\mathrm{KT}} <
%T < T_{\ast}$:
$n\approx 2n_{\mathrm{B}}$ and we have a new
type of metal: a normal bosonic metal. Let us study the
resistivity and specific heat of this metal. At very low
densities, when not only the parameter $|E_{\mathrm{b}}| /Wn
\gg 1$, but also \cite{6}, \cite{7} $\ln \ln
(|E_{\mathrm{b}}| /Wn) {\ \lower-1.2pt\vbox{\hbox{\rlap{$
>
$}\lower5pt\vbox{\hbox{$\sim$}}}}\ }1 $,
the Kosterlitz-Thouless critical temperature is smaller
than the degeneracy temperature  $T_0=${\large
$\frac{nW}{4}$} of a 2D gas of bound pairs. In this
case the bosonic chemical potential at temperatures $T<T_0$
becomes exponentially small and reads:
$\mu_{\mathrm{B}}=-T\exp(-T_0/T)$. As a result, the specific heat
of the system becomes linear \cite{8}: $C_v=nT/T_0$ for
$T<T_0$. Note that since $C_v=n$ for $T>T_0$ there is no
$\lambda$-point in our system.

Finally, the resistivity
%,
%produced by boson-boson interaction
%or scattering on impurities,
changes its behavior \cite{8}
from $R\sim \sqrt{T}$ for $T>T_0$ on
$R\sim\frac{1}{\sqrt{T}} \exp \left( -\frac{T_0}{T}\right)$
for $T < T_0$.

\section{Conclusions}
We would like to emphasize that already such a simple
theory, as presented above, brings a lot of similarities with
the recent results of beautiful tunneling experiments
\cite{9} in underdoped regime of HTSC-materials. These
similarities include a large value of the ratio of {\large
$\frac{2\Delta}{T_c}$}$\sim 8 \div 10$ and a strong
anisotropy of the density of states both in theory and in
experiment. Note that in our theory
$2\Delta/T_c =2 |E_{\mathrm{b}}| /T_{\mathrm{KT}}$.
%
%Moreover, the correlation gap $\Delta$ does not change
%drastically  both in theory and in experiment for
%temperatures above $T_c$.
%
%Of course, there is one substantial difference between our
%calculations and the
%experimental situation in underdoped HTSC-materials: we work in the limit
%$n \rightarrow 0$, while for experimentalists the actual limit is
%$n \rightarrow 1$.

 % Acknowledgements are optional.
\begin{ack}
This work has been supported by the Fonds national suisse de la
recherche scientifique and RFBR grant 98-02-17077.
\end{ack}

% References

\end{document}